\documentclass[twocolumn,aps,showpacs]{revtex4}
\begin{document}
%\def\sqr#1#2{{\vcenter{\hrule height.#2pt\hbox{\vrule width.#2pt
%height#1pt \kern#1pt \vrule width.#2pt}\hrule height.#2pt}}}
%\def\square{\mathchoice\sqr64\sqr64\sqr{4.2}3\sqr{3.0}3}
%\draft

\newcommand{\nd}{\noindent}
\newcommand{\nl}{\newline}
\newcommand{\be}{\begin{equation}}
\newcommand{\ee}{\end{equation}}
\newcommand{\ben}{\begin{eqnarray}}
\newcommand{\een}{\end{eqnarray}}
\newcommand{\nn}{\nonumber \\}
\newcommand{\ii}{\'{\i}}
\newcommand{\pp}{\prime}
\newcommand{\expq}{e_q}
\newcommand{\lnq}{\ln_q}
\newcommand{\quno}{q-1}
\newcommand{\qunoinv}{\frac{1}{q-1}}
\newcommand{\tr}{{\mathrm{Tr}}}

\title{Quantum statistical information contained in a semi-classical Fisher--Husimi measure}

\author{F.~Pennini$^{1,2}$, A.~Plastino$^1$ and G.L.~Ferri$^3$}

%\author{F.~Pennini, A. Plastino\\La Plata National University,
%Argentina  \\and Argentina's National Research Council
%(CONICET)\\C. C. 727,
%1900 La Plata, Argentina\\\\
% G. L. Ferri\\Facultad de Ciencias Exactas, National University La
%Pampa\\Peru y Uruguay, Santa Rosa, La Pampa, Argentina}

\affiliation{$^1$Instituto de F\'{\i}sica La Plata (CONICET-IFLP),
Facultad de Ciencias Exactas, Universidad Nacional de La Plata
(UNLP),
C.C.~727, (1900) La Plata, Argentina\\
$^2$Departamento de F\'{\i}sica, Universidad Cat\'olica del Norte, \\
 Av. Angamos 0610, Antofagasta, Chile\\
$^3$ Facultad de Ciencias Exactas, National University La
Pampa\\Peru y Uruguay, Santa Rosa, La Pampa, Argentina.}

 %\address{
 %$^1$
 %Universidad Nacional de La Plata~(UNLP) and Argentine National
 %Research Council~(CONICET)\\ C.C.~727, 1900 La Plata, Argentina
 %\\
 %$^2$Facultad de Ciencias Exactas y Naturales, La Pampa National
%University} \maketitle

\date{\today}

\begin{abstract}
 \nd We study here the difference between quantum statistical treatments and semi-classical ones, using as
 the main research tool  a semi-classical, shift-invariant Fisher information
 measure built up with Husimi distributions. Its semi-classical
 character notwithstanding, this measure also contains information
 of a purely quantal nature.
 Such a tool allows us to   refine the celebrated Lieb bound for Wehrl entropies and to discover
 thermodynamic-like relations that involve the degree of
 delocalization. Fisher-related thermal uncertainty relations are developed and the degree of
 purity of canonical distributions, regarded as mixed states, is connected  to this Fisher measure as well.


%% 2.50.-r: Probability theory, stochstics processes, and statistics
%% 3.67.-a: Quantum information; 31.15.Gy: Semi-classical methods
%% 05.30.-d: Quantum statistical mechanics

 \end{abstract}

\maketitle

 %%%%%%%%%%%%%%%%%%%%%%%%%%%%%%%%%%%%%%%%%%%%%%%%%%%%%%%%%%%%%%%
\section{Introduction}
 %%%%%%%%%%%%%%%%%%%%%%%%%%%%%%%%%%%%%%%%%%%%%%%%%%%%%%%%%%%%%%%

 \nd A quarter of century before Shannon, R.A.~Fisher advanced a
method to measure the information content of continuous, rather
than digital inputs, using not the binary computer codes but
rather the statistical distribution of classical probability
theory~\cite{roybook,roybook2}. Already in 1980 Wootters pointed
out that Fisher's information measure~(FIM) and quantum mechanics
share a common formalism and both relate probabilities to the
squares of continuous functions~\cite{ThWootters}. Since then,
much  interesting work has been devoted to the manifold physical
FIM--applications.   For
  examples (not an exhaustive list, of course), see, for
  instance,~Refs.~\cite{roybook,Frieden,Renyi,FPPS,Incerteza}).

 \nd Our emphasis in this communication will be placed on
 the study of the differences between (i) statistical treatments of a purely quantal nature,
 on the one hand, and  (ii) semi-classical ones, on the other one. We will show that these differences
 can be neatly expressed entirely in terms of a semi-classical, shift-invariant Fisher
 measure. Additionally, this measure helps to refine the so-called Lieb--bound~\cite{wehrl}
 and connects this refinement with a Fisher-description of
 delocalization in  phase-space, that, of course,
  can be visualized as
information loss. Relations will also be established with an
interesting measure that was early introduced to characterize
delocalization: that of Wehrl's~\cite{wehrl}, for which Lieb
established the above cited lower bound~\cite{Lieb}.

\nd In the wake of a discussion advanced  in Ref.~\cite{Pennini1},
 {\it we will be mainly concerned
 with  building ``Husimi--Fisher'' bridges}. It is well-known that the oldest
 and most elaborate phase-space~(PS) formulation of quantum mechanics
is that of
 Wigner~\cite{wigner,Wlodarz}. To every quantum state a PS function (the
 Wigner one) can be assigned. This PS function can, regrettably enough,
 assume negative values so that a probabilistic interpretation
 becomes questionable. Such limitation was
  overcome  by Husimi~\cite{husimi} (among others). In terms of the concomitant Husimi
probability  distributions,  quantum mechanics can be completely
reformulated~\cite{oconnel,mizrahi}.

 \nd The focus of our
  attention will be, following Ref.~\cite{PRD2753_93}, the  thermal
  description of the harmonic oscillator~(HO) and its
  phase-space delocalization as temperature grows, {\sf in the understanding that
  the HO is, of course,  much more than a mere example}, since nowadays it
is of particular interest for the dynamics of bosonic or fermionic
atoms contained in magnetic traps~\cite{anderson,davis,bradley},
as well as for any system that exhibits an equidistant level
spacing in the vicinity of the ground state, like nuclei or
Luttinger liquids. However, a {\it generalization scheme that
allows for going beyond the HO} will be discussed as well (Cf.
Sect.~\ref{HIOplus}).

\nd The paper is  organized as follows. In Section~\ref{cwehrl} we
briefly review  basic notions  about i) Fisher's information
measure and ii) coherent states and Husimi distributions for a
system's thermal state.  The nucleus of this communication is
developed in Section~\ref{deloca}: appropriately employing the
semi-classical, shift-invariant Fisher measure so as to uncover
the rather surprising amount of purely quantum information that it
carries. Thermodynamic-like relations are derived in
Section~\ref{thermo}. A survey on thermal uncertainty relations is
introduced in Section~\ref{uncerthermal} in order to show that,
via Fisher, quantum and semi-classical uncertainties can be
linked. We are able to establish some special connections between
degrees of purity in Section~\ref{purities1}, and to go beyond the
harmonic oscillator in~Section~\ref{HIOplus}. Finally, some
conclusions are drawn in Section~\ref{conclusions}.

%%%%%%%%%%%%%%%%%%%%%%%%%%%%%%%%%%%%%%%%%%%%%%%%%%%%%%%%
  \section{Background material}
   \label{cwehrl}
%%%%%%%%%%%%%%%%%%%%%%%%%%%%%%%%%%%%%%%%%%%%%%%%%%%%%%%%%%

%%%%%%%%%%%%%%%%%%%%%%%%%%%%%%%%%%%%%%%%%%%%%%%%%%%%%%%%%%%%%
\subsection{Fisher's information measure}
  %%%%%%%%%%%%%%%%%%%%%%%%%%%%%%%%%%%%%%%%%%%%%%%%%%%%%%%%%%%%%
  \label{Fisher}
\vskip 3mm
  \noindent  One important information measure is that advanced by
R.~A.
  Fisher in the twenties (for a detailed study
see Ref.~\cite{roybook,Frieden}).
  Consider a $[\theta\,-\,{\bf x}]$ ``scenario" in which we deal
  with a system specified by a physical
  parameter $\theta$,  while ${\bf x}$ is a stochastic variable $({\bf
  x}\,\in\,\Re^{N})$,
  and
  $f_\theta({\bf x})$ the probability density for ${\bf x}$
  (that depends also on  $\theta$).  One makes a
  measurement of
   ${\bf x}$ and
  has to best infer $\theta$ from this  measurement,
   calling the
    resulting estimate $\tilde \theta=\tilde \theta({\bf x})$.
How well $\theta$ can be determined? Estimation
theory~\cite{cramer}
   states that {\sf the best possible estimator} $\tilde
   \theta({\bf x})$, after a very large number of ${\bf x}$-samples
  is examined, suffers a mean-square error $e^2$ from $\theta$ that
  obeys a relationship involving Fisher's $I$, namely, $Ie^2=1$,
  where the Fisher information measure~$I$ is of the form

  \ben
  I(\theta)&=&\int \,\mathrm{d}{\bf x}\,f_\theta({\bf
  x})\left\{\frac{\partial \ln f_\theta({\bf x})}{
  \partial \theta}\right\}^2,\,\,\,{\rm i.e.},
  \cr I(\theta) &\equiv& \left\langle \left\{\frac{\partial \ln f_\theta({\bf x})}{
  \partial \theta}\right\}^2\right\rangle_{f}  \label{ifisher}.
  \een
  \noindent  This ``best'' estimator is the so-called {\it efficient}
estimator.
  Any other estimator exhibits a larger mean-square error. The only
  caveat to the above result is that all estimators be unbiased,
  i.e., satisfy $ \langle \tilde \theta({\bf x}) \rangle=\,\theta
  \label{unbias}$.  Fisher's information measure has a lower bound:
 no matter what parameter of the system  one chooses to
  measure, $I$ has to be larger or equal than the inverse of the
  mean-square error associated with  the concomitant   experiment.
  This result, \be I\,e^2\,\ge \,1,\ee is referred to as the
  Cramer--Rao bound~\cite{roybook}.

 \subsubsection{Shift invariant $I-$measure} \nd A particular $I$-case is of great importance:
 that of translation families~\cite{roybook,Renyi},
  i.e., distribution functions~(DF) whose {\it
   form} does not change under $\theta$-displacements. These~DF
   are shift-invariant (\`a la Mach, no absolute origin for
  $\theta$), and for them
   Fisher's information measure adopts the somewhat simpler
   appearance~\cite{roybook}

   \be\label{shift}
  %I_{\tau}= \left\langle \left\{\frac{\partial \ln
  %f({\bf x})}{
  %\partial {\bf x}}\right\}^2 \right\rangle_f = \int \mathrm{d}{\bf x}\,f({\bf
%x})\left\{\frac{\partial \ln
%  f({\bf x})}{
 % \partial {\bf x}}\right\}^2.
 I_\tau=\int \,\mathrm{d}{\bf x}\,[1/f({\bf x})]\,\left\{\nabla f({\bf x})\right\}^2.
   \ee
\noindent The shift-invariant  measure $I_\tau$ as a function of
phase-space coordinates $\tau \equiv (x,\,p)$ will be the
protagonist of this communication. Notice that $I_\tau$   is an
{\it additive} measure~\cite{roybook}. Thus, since $x\,\,{\rm
  and}\,\, p$  are independent  variables, we will have  $I_\tau(x+p)=I_\tau(x)+I_\tau(p)$.
    We postpone writing
 down the pertinent Fisher expression  until after having briefly reviewed
 coherent states [see Eq.~(\ref{II}) below].

%%%%%%%%%%%%%%%%%%%%%%%%%%%%%%%%%%%%%%%%%%%%%%%%%%%%%%%%%%
\subsection{ Coherent states and Husimi distribution}
  %%%%%%%%%%%%%%%%%%%%%%%%%%%%%%%%%%%%%%%%%%%%%%%%%%%%%%%%

\nd The semi-classical Wehrl entropy $W$ is a  useful measure of
localization in phase-space ~\cite{wehrl,Gnutzmann}. It is  built
up using coherent states $\vert
z\rangle$~\cite{wehrl,PRD2753_93,Glauber} and
  constitutes a  powerful  tool in statistical physics.
 Of course, coherent states are eigenstates of a general
 annihilation operator $a$, appropriate for the problem at hand,
 i.e., $ a\vert z \rangle=z\vert z
\rangle$~\cite{Glauber,klauder,Schnack}.   The pertinent
$W-$definition reads

\be
    W=-\int \frac{\mathrm{d}x\, \mathrm{d}p}{2 \pi \hbar}
 \mu(x,p)\, \ln
    \mu(x,p),\label{i1}\ee clearly a Shannon-like measure~\cite{katz}
     to which~MxEnt considerations can be applied (see
    Sect. \ref{delorevisi} below).
\vskip 1mm \nd      The functions $\mu$, commonly referred to as
Husimi   distributions~\cite{husimi}, are the diagonal elements of
the density matrix in the coherent-state
      basis $\vert z \rangle$, i.e., \be \mu(x,p)=\langle z|
      \rho|z\rangle.\ee They  are  ``semi-classical'' phase-space distribution
 functions  associated  to the  density matrix~$\rho$ of the
 system~\cite{Glauber,klauder,Schnack}.  The
 distribution  $\mu(x,p)$ is    normalized in the fashion

 \be \label{lanorma}
\int \frac{\mathrm{d}x\, \mathrm{d}p}{2 \pi \hbar}\,
    \mu(x,p)= 1.
   \ee
    Indeed,   $\mu(x,p)$ is a
 Wigner--distribution $D_W$, smeared over an $\hbar$ sized region of
phase-space~\cite{PRD2753_93}.
   The smearing renders $\mu(x,p)$ a positive function, even if $D_W$
  does not have
    such a character. The semi-classical Husimi probability
distribution
 refers to a {\it special type} of probability:
  that for simultaneous but approximate location of position and
 momentum in phase space~\cite{PRD2753_93}.
The uncertainty principle manifests itself  through the inequality
\be \label{UPHO} W \geq 1, \ee which was first conjectured by
Wehrl~\cite{wehrl} and later proved by Lieb~\cite{Lieb}. Equality
holds if $\rho$ is a coherent state~\cite{wehrl,Lieb}.

\vskip 2mm \nd  For dealing with equilibrium states at a
temperature $T$ in statistical mechanics one usually  represents
the  system's state as an incoherent superposition ({\it mixed
state}) of eigenenergies~$E_n$ weighted by the exponential
Boltzmann factor $\exp{(-\beta E_n)}$, with $\beta=1/kT$    the
inverse temperature and, and $k$ the Boltzmann constant (we take
$k=1$ hereafter). In other words, use is made of  Gibbs's
canonical distribution, whose associated, ``thermal'' density
matrix is given by

  \be \label{canonic} \rho=Z^{-1}e^{-\beta
 \mathcal{H}}, \ee
    with $Z=Tr(e^{-\beta \mathcal{H}})$  the partition function.
    In order to conveniently write down an expression for
    $W$ one considers, for the Hamiltonian $\mathcal{H}$,
    its eigenstates   $|n\rangle$ and eigenenergies $E_n$, because
 one can
 always write~\cite{PRD2753_93}

 \be \mu(x,p)=\langle z| \rho|z \rangle=\frac{1}{Z}\, \sum_{n}
 e^{-\beta
   E_n}\,|\langle z|n\rangle|^2,\label{husimi}
 \ee with
\be \label{husimitraza} \tr\,\rho \equiv \int \frac{\mathrm{d}x\, \mathrm{d}p}{2 \pi \hbar}
   \, \mu(x,p) =1.\ee
 A useful route to $W$ starts then with Eq.~(\ref{husimi}) and
 continues with Eq.~(\ref{i1}).
  Quantum-mechanical phase-space distributions expressed in terms
of the  coherent states $\vert z \rangle$ of the harmonic
oscillator have proved to be   useful in different
contexts~\cite{Glauber,klauder,Schnack}.   Particular reference is
to be made to  the illuminating work
 of Andersen and Halliwell~\cite{PRD2753_93}, who discuss, among other
 things, the concepts of Husimi distributions and Wehrl entropy.

%%%%%%%%%%%%%%%%%%%%%%%%%%%%%%%%%%%%%%%%%%%%%%%%%%%%%%
\subsection{Harmonic oscillator Husimi results}
%%%%%%%%%%%%%%%%%%%%%%%%%%%%%%%%%%%%%%%%%%%%%%%%%%%%%%%
 \nd The above considerations are of a general character. In the
special case of the
 harmonic oscillator, whose Hamiltonian has the form
 \be \label{HOH} \mathcal{H}_{HO} =
\hbar \omega \,
    [ a^{\dagger}  a +1/2]= (\hbar \omega/2) \, [ a^{\dagger}  a + a  a^{\dagger}],\ee
    the (complex) {\it eigenvalues} $z$ of
 the annihilation operator $a$ are given by

    \be \label{z}
 z=
\frac{1}{2}\,\left(\frac{x}{\sigma_x} +i
\frac{p}{\sigma_p}\right),
    \ee
 where the variables $x$ and $p$,
 are
  scaled by their respective variances ($\sigma$) for the HO ground
state  $$\sigma_x =(\hbar/2m\omega)^{1/2};\,\,\, \sigma_p=(\hbar m
\omega/2)^{1/2};\,\,\, \sigma_x \sigma_p =\hbar/2.$$ The Husimi
distribution $\mu(x,p)$ adopts the
appearance~\cite{Pennini1,PRD2753_93}

\be \mu(x,p)\equiv \mu(z)=(1-e^{-\beta \hbar
\omega})\,e^{-(1-e^{-\beta \hbar \omega})|z|^2},\label{cstate} \ee
which is normalized in the fashion

\be \int \frac{\mathrm{d}^2 z}{\pi}\,\mu(z)=1, \ee where
$\mathrm{d}^2 z/\pi= \mathrm{d}(Re\, z)\mathrm{d}(Im\, z)
=\mathrm{d}x\mathrm{d}p/(2\pi\hbar)$ is the differential element
of area in the $z$ plane~\cite{Glauber} and $z$ is given by~Eq.
(\ref{z}). The HO-Wehrl's measure becomes then~\cite{Pennini1}

\be \label{Wehrl} W=1-\ln{[1-e^{-\beta\hbar\omega}]}.\ee
%%%%%%%%%%%%%%%%%%%%%%%%%%%%%%%%%%%%%%%%%%%
\subsubsection{Semi-classical purity}
%%%%%%%%%%%%%%%%%%%%%%%%%%%%%%%%%%%%%%%%%%
\nd The degree of purity of a density matrix $\rho$ is given by
$Tr \,\rho^2$~\cite{ufano}. Its inverse, the so-called {\it
participation ratio}

 \be \label{participa} R(\rho)=\frac{1}{\tr \rho^2}, \ee  is particularly
convenient for calculations~\cite{casas}. It varies from unity for
pure states to N for totally mixed states~\cite{casas}. It may be
interpreted as the effective number of pure states that enter a
quantum  mixture. Here we will consider the ``degree of purity"
$d_\mu$ of a semi-classical distribution, given by

\be
\label{mixto} d_\mu= \int \frac{\mathrm{d^2}z}{
   \pi} \, \mu^2(z) \le 1,  \ee
to be evaluated below [Cf.~(\ref{purito})].

%%%%%%%%%%%%%%%%%%%%%%%%%%%%%%%%%%%%
\subsubsection{Quantal purity}
%%%%%%%%%%%%%%%%%%%%%%%%%%%%%%%%%%%%%%
\nd For the quantum mixed HO-state

\be  \rho_{HO}=e^{-\beta \mathcal{H}_{HO}}/ Z, \label{rho} \ee
with $Z=e^{\beta\hbar\omega/2}/(e^{\beta\hbar\omega}-1)$ the
partition function~\cite{pathria1993},  we have a degree of purity
$d_{\rho}$ given by (see the detailed study by Dodonov~\cite{dodonov})

 \be d_{\rho}=\frac{e^{-\beta\hbar\omega}}{Z^2}\sum_{n=0}^{\infty}\,e^{-2\beta\hbar\omega
n} \label{puroQ}, \ee leading to

\be d_{\rho}=\tanh (\beta \hbar\omega/2), \label{Qpuro} \ee
where $0 \leq d_{\rho}\leq 1$.
 Thus,
Heisenberg' uncertainty relation can be cast in the fashion

\be \Delta_ x\,\Delta
p=\frac{\hbar}{2}\,\coth(\beta\hbar\omega/2), \label{dodono} \ee
where $\Delta x$ and $\Delta p$ are the quantum variances for the
canonically conjugated observables $x$ and $p$~\cite{dodonov}

 \be \label{delta1} \Delta x\,\Delta
p=\frac{\hbar}{2}\,\frac{1}{d_{\rho}}, \ee which is to be compared
to the semi-classical result that we will derive below [Cf.~(\ref{virial})], on the basis of Eq.~(\ref{mixto}).

%%%%%%%%%%%%%%%%%%%%%%%%%%%%%%%%%%%
\subsubsection{Basic quantal HO relations}
%%%%%%%%%%%%%%%%%%%%%%%%%%%%%%%%%%
 \nd We write down now, for
future reference,
  well-known {\it quantal} HO-expressions for,
  respectively, the entropy~$S$, the mean energy $U$, the
  mean excitation energy~$E$,
  and the specific heat~$C$~\cite{pathria1993}

   \ben \label{textbook} S&=& \beta \frac{\hbar \omega}{e^{\beta \hbar
\omega}-1}-
  \ln{\{1-e^{-\beta \hbar
  \omega}\}},\nonumber\\
  U&=&\frac{\hbar\omega}{2}+E=
\left[\frac{\hbar\omega}{2}+\frac{\hbar\omega}{e^{\beta\hbar\omega}-1}\right],\\
  C&=& -\beta^2\left(\partial U/\partial
  \beta\right)=\left[\frac{\hbar\omega\beta}{e^{\beta\hbar\omega}-1}\right]^2\,e^{\beta\hbar\omega}\nonumber.
 \een

%%%%%%%%%%%%%%%%%%%%%%%%%%%%%%%%%%%%%%%%%%%%%%%%%%%%%%%%%%%%%%%
\section{Semi-classical Fisher's measure} \label{deloca}
%%%%%%%%%%%%%%%%%%%%%%%%%%%%%%%%%%%%%%%%%%%%%%%%%%%%%%%%%%%%%%%%

\nd According to the preceding considerations, the Fisher
  measure (\ref{shift})  associated to the probability distribution
$\mu(x,p)\equiv\mu(\tau)$ will
  be of the form~\cite{Renyi}

  \be  I_{\tau}=\int \frac{\mathrm{d}^2 z}{\pi}\mu(z) \left\{\sigma_x^2\left[\frac{\partial \ln
  \mu(z)}{\partial x}\right]^2 +\sigma_p^2\left[\frac{\partial \ln
  \mu(z)}{\partial p}\right]^2\right\}.\label{II}
   \ee

\nd In the HO instance  $I_{\tau}$ becomes, given the
$\mu-$expression ~(\ref{cstate}) \cite{Pennini1},

  \be\label{IF}
  I_{\tau}=1-e^{-\beta \hbar \omega}.\ee Since the temperature lies between zero
and infinity, the range of values of
  $I_\tau$ is \be \label{rango} 0 \le I_\tau \le 1. \ee
The above entails, via Eq.~(\ref{textbook}), that the {\it
quantal} HO expressions for $E$ and $S$ can be expressed in terms
of the {\it semi-classical} measure $I_{\tau}$ \ben
\label{Fishtextbook} \frac{E}{\hbar \omega}&=& \frac{ 1-I_{\tau}}{
I_{\tau}}\cr S&=& \beta \hbar \omega \frac{ 1-I_{\tau}}{ I_{\tau}}
- \ln{I_\tau},\een which shows that the semi-classical, Husimi
based $I_\tau-$information measure does contain purely
quantum-statistical information.
 The fact that the quantal HO-thermodynamic quantities (q.t.q.) $U=
 \hbar\omega/2+E$ and $S$ can be entirely expressed in terms of
 $I_\tau$, implies that, a posteriori,   all q.t.q.'s can be written in these terms.
  We emphasize thus, as a {\sf new result (the first of this communication)}, the fact that
  the semi-classical quantity  $I_\tau$ contains all the relevant  HO-statistical quantum
  information.

\nd In the wake of Eq.~(\ref{Wehrl}), for the
 equilibrium thermal state $\rho$, the entropy $W$ and the shift
invariant Fisher measure $I_\tau$ are related according
to~\cite{Pennini1}

 \be W+\ln I_\tau=1,   \label{fla}\ee
\nd i.e., \be \label{nueva}  e^{W}\,\,I_\tau=e,\ee
 so that
the two measures become
 complementary informational quantities, which entails that
 the quantal expressions for $E$ and $S$ can
also be cast in terms of $W$.

\nd How stable are the results (\ref{Fishtextbook})? To answer
this question we resort to a numerical procedure. We build a
two-dimensional lattice large enough to accommodate numerical
integration in phase-space with  precision $1$ in $10^{-16}$ and
vary, at each lattice-point, $\mu(z)$ by a random, small amount
$\delta \mu(z)= \xi R_{nd}\mu(z)$, with $R_{nd}$ random and $\xi$
a small quantity. We then evaluate the concomitant $\delta W$ and
find $\delta W \propto \xi^2$. Thus, first-order changes in
$\mu(z)$ lead to second-order changes in $\delta W$, and,
consequently, in $\ln{I_\tau}$. The results (\ref{Fishtextbook})
are stable against small changes in the Husimi~PD.
%%%%%%%%%%%%%%%%%%%%%%%%%%%%%%%%%%%%%%%
\subsection{MaxEnt approach}
%%%%%%%%%%%%%%%%%%%%%%%%%%%%%%%%%%%%%%%
\nd Note also that, from Eq.~(\ref{cstate}), we can conveniently
recast the HO-expression for $\mu$ into the Gaussian fashion

 \be
\label{minueva} \mu(z)= I_\tau\,e^{-I_\tau\,\vert z\vert^2}, \ee
peaked at the origin. The Fisher measure $I_\tau$ of
Eq.~(\ref{minueva}) is clearly of the maximum entropy
(MaxEnt)~\cite{katz} form (compare, for instance, with Eq.~(4.2)
of~Ref.~\cite{canosa}). As a consequence, it can be viewed in the
following light. Assume we know a priori the value
$\mathcal{E}_\nu=\langle \hbar \omega \vert z\vert^2\rangle_\nu$.
We wish to determine the distribution $\nu(z)$ that maximizes the
Wehrl entropy $W$ under this $\mathcal{E}_\nu-$value constraint.
Accordingly, the MaxEnt distribution will be~\cite{katz}

\be \label{maxent} \nu(z) = e^{-\lambda_o}e^{-\eta\,E(z)},\ee with
$\lambda_o$ the normalization Lagrange multiplier and $\eta$ the
one associated to $\mathcal{E}_\nu$. According to MaxEnt tenets we
have~\cite{katz}

\be \label{lambdacero} \lambda_o=\lambda_o(\eta)=
\ln{\int\,\frac{\mathrm{d}^2 z}{\pi}\,e^{-\eta \,\,\hbar \omega
\vert z\vert^2}}=-\ln{(\eta \,\,\hbar \omega)}.\ee Now, the
$\eta-$multiplier is determined by the relation~\cite{katz}

\be \label{eta} -\mathcal{E}_\nu = \frac{\partial
\lambda_o}{\partial \eta}=-\frac{1}{\eta}.  \ee If we choose the
Fisher-Husimi constraint given by Eq. (\ref{mH}) $\mathcal{E}_\mu=
\hbar
  \omega/I_\tau$, this results in

  \ben \label{lagrange}  \eta &=& I_\tau/(\hbar \omega),\,\,\,{\rm and,\,\,\,from\,\,\, (\ref{lambdacero})}
  \cr  \lambda_o&=& -\ln{I_\tau}\,\,\,{\rm i.e.,}   \cr
  e^{-\lambda_o}&=& I_\tau\,\,\,{\rm and}  \cr
   \nu(z) &=&  I_\tau  \,  e^{-I_\tau\,\vert z\vert^2}  \equiv \mu(z).\een
We have thus shown that the HO-Husimi distributions  are
MaxEnt-ones with the  semi-classical excitation energy ~(\ref{mH})
as a constraint, {\sf our 2th new result}. It is clear from
Eq.~(\ref{lagrange}) that $I_\tau$ plays there the role of an
``inverse temperature". This means that we can think of a quantity
$T_W$ associated to the Wehrl measure on account of\be
\label{wehrlT}  \mu(z) = I_\tau  \, e^{-(I_\tau/\hbar
\omega)\,\hbar \omega \vert z\vert^2}=  I_\tau \,
e^{-\mathcal{E}_\mu/T_W}, \ee which entails \be \label{TWD}
T_W=\hbar \omega/ I_\tau;\,\,\,(\hbar \omega \le T_W \le \infty).
\ee Due to the semi-classical nature of both $W$ and $\mu$, $T_W$
has a lower bound greater than zero. At this stage we introduce a
``delocalization factor" $D$ \ben & D= T_W/\hbar
\omega\,\,\Rightarrow \cr & W= 1+\ln{T_W}-\ln{\hbar
\omega}=1+\ln{D}, \label{newD} \een to be discussed next.

\subsection{Delocalization revisited}
\label{delorevisi} \nd As stressed above, $W$ has been conceived
as a delocalization measure. The preceding considerations clearly
motivate one to regard, as well,  the Fisher measure built up with
Husimi distributions $\vert z\vert$~(Cf.~Eq.~(\ref{z})) as a
``localization estimator" in phase space.  It has been shown
already in~Ref.~\cite{Pennini1} that efficient estimation (meaning
that the Cramer--Rao lowest bound is reached) is
 possible for all temperatures $T$.
The HO-Gaussian expression for $\mu$ (\ref{minueva}) illuminates
the fact  that the Fisher measure controls height, on the one
hand, and  spread, on the other one (that is $\sim
[2I_\tau]^{-1}$). Obviously,  {\sf spread is here a ``phase-space
delocalization indicator".} This fact is reflected by the quantity
$D$ introduced above.

\nd Thus, an  original physical interpretation of Fisher's measure
emerges: {\sf localization control}.  The inverse of the Fisher
measure, $D$, {\it turns out then to be a
delocalization-indicator}. Notice also that

 \be \label{dIdT}
\frac{\mathrm{d}I_\tau}{\mathrm{d}T} =-\frac{\hbar\omega}{T^2}\,
e^{-\beta\hbar\omega},\ee so that Fisher's information decreases
exponentially as the temperature grows. Our Gaussian distribution
loses phase-space ``localization" as energy and/or temperature are
injected into our system, as reflected via $T_W$ or $D$. Remark
that~(\ref{newD}) {\it complements the Lieb bound}~$W\ge 1$. $W$
 exceeds unity by virtue of delocalization effects, and this can
be expressed using the shift-invariant Fisher measure {\sf (our
3rd. new result)}. We will now show that $D$ is proportional to
the system's energy fluctuations.

%%%%%%%%%%%%%%%%%%%%%%%%%%%%%%%%%%%%%%%%%%%%%%%%%%%%%%%%%%%%%%%
\subsection{Second moment of the Husimi distribution}
 %%%%%%%%%%%%%%%%%%%%%%%%%%%%%%%%%%%%%%%%%%%%%%%%%%%%%%%%%%%%%%%
\label{moment} \nd The second moment of the Husimi distribution
$\mu(z)$ given by~Eq. (\ref{minueva}) is defined as~\cite{sugita}

\be M_2=\int \frac{\mathrm{d}^2 z}{\pi}\,\mu^2(z), \ee
 that, after explicit
evaluation of $M_2$ reads
 \be M_2=\frac{I_\tau}{2}. \ee
 Using now (\ref{newD})  we
conclude that \be M_2(D)=\frac{1}{2\, D}. \ee In order to give
physical meaning to this result consider the
  energy-fluctuations evaluated via the distribution function
$\mu(z)$. The semi-classical energy $\mathcal{E}_\mu$~is
 \be
\label{Esemi1} \mathcal{E}_\mu= \int \frac{\mathrm{d}^2 z}{\pi
}\,\,\mu(z)\,E(z), \ee with~\cite{Pennini2}

\be \label{Esemi2} E(z)= \langle z\vert H \vert z\rangle - \frac{\hbar
\omega}{2} = \hbar \omega \vert z \vert^2 =  \langle z\vert \hbar
\omega   a^\dag a \vert z\rangle. \ee
   Thus,

   \be \mathcal{E}_\mu=\frac{\hbar
  \omega}{I_\tau} \label{mH}. \ee Comparing now with~Eq.~(\ref{textbook})    we see that
  \be \label{energdiff} \mathcal{E}_\mu - E= \hbar\omega, \ee
  i.e., the difference between the semi-classical energy $\mathcal{E}_\mu$ and the
  mean quantum   energy  $U$ equals $\hbar \omega/2$, the
  vacuum-energy, independently of $T$.
   For our purposes we need also the semi-classical mean value of $(\mathcal{E}^2)_\mu$
 \be (\mathcal{E}^2)_\mu=\int \frac{\mathrm{d}^2 z}{\pi}\, \mu(z)\,E(z)^2,
 \ee i.e.,
  \be (\mathcal{E}^2)_\mu=2\,\left(\frac{ \hbar
  \omega}{I_{\tau}}\right)^2, \ee so that, finally,
  our energy-fluctuations turn out to be

 \be \Delta_{\mu}
  E=\frac{\hbar \omega}{I_{\tau}}=\hbar \omega\,D, \ee
with $(\Delta_{\mu} E)^2=(\mathcal{E}^2)_\mu-\mathcal{E} _\mu^2$.
As a consequence, we get

\be D=\frac{\Delta_{\mu} E}{\hbar \omega}. \ee {\sf An important
new result (our 4th. one)} is thus obtained:  the delocalization
factor $D$ represents
 energy-fluctuations expressed in $\hbar \omega-$terms. {\sf
Delocalization is clearly  seen to be the counterpart of energy
fluctuations!}

%%%%%%%%%%%%%%%%%%%%%%%%%%%%%%%%%%%%%%%%
\subsection{Purity and delocalization}
%%%%%%%%%%%%%%%%%%%%%%%%%%%%%%%%%%%%%%%%%%%
\nd Uncertainty  arguments can also be used for  ``purity"
purposes~\cite{ufano}~(Cf. (\ref{mixto})). The semi-classical
degree of purity is given by

 \be d_\mu=\int \frac{\mathrm{d}^2
z}{\pi}\,\mu^2(z)=\frac{I_\tau}{2}=\frac{1}{2D}, \label{purito}\ee %where
 so that $0\leq d_\mu \leq 1/2$. Notice that $I_\tau$ is [Cf. Eq.
 (\ref{shift})] is a mean value evaluated with a semi-classical
 distribution, which helps to achieve a first understanding of the low value (1/2) of
 the degree of purity's upper bound. A more complete discussion is
 given below.

%%%%%%%%%%%%%%%%%%%%%%%%%%%%%%%%%%%%%%%%%%%%%%%%%%%%%%%%%%%%%%%%%%
\section{Thermodynamics-like relations}
%%%%%%%%%%%%%%%%%%%%%%%%%%%%%%%%%%%%%%%%%%%%%%%%%%%%%%%%%%%%%%%%
\label{thermo} \nd We now go back to Eq.~(\ref{textbook}) and take
a hard look at the entropic expression and see that we can recast
the entropy $S$ in terms of the quantal mean excitation energy $E$
and the delocalization factor $D$ (the information $I_\tau$)~as

\be \label{newentro} \frac{E}{T}= S- \ln{D},\label{thermal0} \ee
i.e., if one injects into the system some  excitation energy $E$,
expressed in  ``natural" $T$ units (remember that we have set
$k=1$), it is apportioned partly as heat dissipation via $S$ and
partly via delocalization. More precisely, the part of this energy
not dissipated is that employed to delocalize the system in phase
space. Now, since $W=1-\ln{I_\tau}=1+\ln{D}$, the above equation
can be recast in alternative forms, as

 \be
\label{thermalx} S= \frac{E}{T}+\ln{D}= \frac{E}{T}- \ln{I_\tau};
\,\,\,\,{\rm or:} \ee \be \label{thermal1} W = 1+S -\frac{E}{T},
\ee implying

\be \label{thermal3}  W-S \mapsto 0\,\,\,\,{\rm
for}\,\,\,\,T\mapsto \infty, \ee which is a physically sensible
one and \be W-S \,\,\,\,\mapsto 1\,\,{\rm for}\,\,T\mapsto
0,\label{thermal4} \ee as it should, since $S=0$ at $T=0$ (third
law of thermodynamics), while $W$ attains there its Lieb's lower
bound of unity.

One finds in Eq.~(\ref{thermalx})  some degree of resemblance to
thermodynamics's fist law. To reassure
 ourselves on this point,
 we slightly changed our underlying canonical probabilities (\ref{canonic}), multiplying them by a
 factor $\mathcal{F}={\rm random\,\, number} / 100$, i.e., we
 generated random numbers according to the normal distribution
 divided by $100$ to obtain the above factors. This entails new
``perturbed"  probabilities $P_i$, conveniently normalized ($ \sum
P_i  = 1$).  With them we evaluate the concomitant changes $dS$,
$dE$ (we did this 50 times, with different random numbers in each
instance) and verified that, numerically,  the difference
$dS-\beta dE \sim 0$. The concomitant results are plotted in
Fig.~\ref{F1}. Since, as stated, numerically $dS = (1/T)\, dE$,
this entails, from Eq.~(\ref{thermalx}), $dI_\tau / I_\tau \simeq
0$. The physical connotations are as follows: if the {\it only}
modification effected is that of a ``population" change $\delta
p_i$ in the $p_i$, this implies that the system undergoes a heat
transfer process~\cite{pathria1993} for which thermodynamics'
first law implies $dU= TdS$ and this is numerically confirmed in
the plots of  Fig.~\ref{F1}.
 The null contribution of $\ln{I_\tau}$ to this process  suggests  that
delocalization (not a thermodynamic effect, but a dynamic one) can
be looked at as behaving (thermodynamically) like a kind of
``work''.

\vskip 2mm

\nd  Now, since (a) [Cf.~(\ref{IF})] $ I_{\tau}=1-e^{-\beta \hbar
\omega},$ and (b) the mean energy of excitation is $E= \hbar
\omega/[\exp{(\beta \hbar\omega)}-1]$ one also finds, for the
quantum-semi-classical difference (QsCD) $S-W$ the original (as
far as we know) result  \ben & W-S= 1-[(I_\tau-1)/I_\tau]
\ln(1-I_\tau)=F_1(I_\tau) \label{thermal2}\een \nd Moreover, since
$0\le F_1(I_\tau) \le 1$, we see that, always, $W \ge S$, as
expected, since the semi-classical treatment contains less
information than the quantal one. Note that the QsCD can be
expressed exclusively on Fisher's information measure. This is,
the quantum-semi-classical entropic difference $S-W$ may be given
in $I_\tau-$terms only, which is a new Fisher-property {\sf (our
5th. new result)}. Fig.~\ref{F2} depicts $S$, $\beta E$, and
$\ln{D}$ vs. the dimensionless quantity $t=T/\hbar \omega$.
According to Eq.~(\ref{newentro}), entropy is apportioned in such
a way that
\begin{itemize} \item part of it originates from excitation energy
and \item the remaining is accounted for by phase-space
delocalization. \end{itemize} A bit of algebra allows one now to
express the rate of entropic change per unit temperature increase
as

\be \label{rate} \frac{\mathrm{d}S}{\mathrm{d}T}=
\beta\,\frac{\mathrm{d}E}{\mathrm{d}T}= \beta\,C= \hbar\omega\,
\frac{1}{T}\,\frac{\mathrm{d}D}{\mathrm{d}T}, \ee entailing  \be
\label{calor} C= \hbar\omega\, \frac{\mathrm{d}D}{\mathrm{d}T}.
\ee

\nd In the case of the one dimensional HO we see that  {\sf the
specific heat measures delocalization change per unit temperature
increase}. Also, $dE/dT \propto dD/dT$, providing us with a very
simple relationship between (mean) excitation energy changes and
delocalization ones {\sf (our 6th new result)} \be \label{relata}
\frac{\mathrm{d}E}{\mathrm{d}D}= \hbar \omega.\ee

%%%%%%%%%%%%%%%%%%%%%%%%%%%%%%%%%%%%%%%%%%%%%
\section{ ``Thermal" uncertainties}
%%%%%%%%%%%%%%%%%%%%%%%%%%%%%%%%%%%%%%%%%%%%%
\label{uncerthermal}
\nd  Thermal uncertainties express the effect of
temperature on Heisenberg's celebrated relations (see, for
instance~\cite{Incerteza,mandelbrot,flang,dodonov}). We use now a
result obtained in  Ref.~\cite{PRD2753_93} (equation (3.12)),
where the authors cast Wehrl's information measure in terms  of
the ``coordinates"'s variances $\Delta_{\mu}x$ and
$\Delta_{\mu}p$, obtaining

 \be \label{Wentro}
W=\ln\left\{\frac{e}{\hbar}\,\Delta_{\mu}x\,\Delta_{\mu}p\right\}.
\ee In the present context,  the relation   $W=1-\ln I_\tau$
allows us to conclude that~\cite{Pennini1}

 \be \label{nuevaincert} I_{\tau}\,
\Delta_{\mu} x \,\Delta_{\mu} p= \hbar, \label{un} \ee {\sf which
can be regarded as a ``Fisher uncertainty principle"} and adds
still another meaning to $I_\tau$: since, necessarily,
$\Delta_{\mu} x \,\Delta_{\mu} p \ge
 \hbar/2$, it is clear that $I_{\tau}/2$ is the ``correcting
 factor" that permits one to reach  the uncertainty's  lower bound
 $\hbar/2$.

\nd As stated above, phase space ``localization" is possible, with
Husimi distributions, only up to~$\hbar$. This is to be compared
to the uncertainties evaluated in a purely quantal fashion,
without using Husimi distributions. By recourse to the virial
theorem~\cite{pathria1993} one easily ascertains
that~\cite{Pennini1} \ben  \Delta x\,\Delta
p&=&\frac{\hbar}{2d_{\rho}}=\frac{\hbar}{2}\,\frac{e^{\beta \hbar
\omega}+1}{e^{\beta \hbar \omega}-1}\,\Rightarrow \cr \cr \cr
\Delta_{\mu}x\,\Delta_{\mu}p&=&\frac{2\,\Delta x\,\Delta
p}{1+e^{-\beta \hbar \omega}}. \label{virial} \een

\nd As $\beta \rightarrow \infty$, $\Delta_\mu \equiv
\Delta_{\mu}x\,\Delta_{\mu}p$ is twice the minimum quantum value
for $\Delta x \Delta p$, and $\Delta_\mu \rightarrow \hbar$, the
``minimal" phase-space cell. The quantum and semi-classical
results do coincide at very high temperature, though. Indeed, one
readily verifies~\cite{Pennini1} that
 Heisenberg's uncertainty relation, as a function of both
frequency and temperature, is governed by a thermal ``uncertainty
function" $F$ that  acquires the aspect

 \be  F(\beta,\omega)=\Delta x\,\,
\Delta p=\frac{1}{2}\left[\Delta_\mu+\frac{E}{\omega}\right]. \ee

Within the present context $F$ can be recast as

\be
F(\beta,\omega)=\frac12\left[\hbar\, D +\frac{E}{\omega}\right],
\label{R7} \ee  so that, for $T$ varying in $[0,\infty]$, the
range of possible $ \Delta x\,\, \Delta p$-values is
$[\hbar/2,\infty].$ Eq.~(\ref{R7}) is a ``Heisenberg--Fisher"
thermal uncertainty~(TU) relation (for a discussion of the~TU
concept see, for instance,~\cite{Incerteza,mandelbrot,flang}).
$F(\beta,\omega)$ grows with both  $E$ and $D$. The usual result
$\hbar/2$ is attained for minimum $D$ and zero excitation energy.
As for $dF/dT$,  one is able to set $F\equiv F(E,D)$, since
$2dF=\hbar dD+ \omega^{-1} dE$. Remarkably enough, the two
contributions to $dF/dT$ are easily seen to be equal and $dF/dT
\rightarrow (1/\omega)\,\,{\rm for}\,\, T \rightarrow \infty$. One
can also write \be \label{uncer} \left(\frac{\partial F}{\partial
D}\right)_E = \frac{\hbar}{2}; \,\,\, \left(\frac{\partial
F}{\partial E}\right)_D = \frac{1}{2\omega}, \ee providing us with
a thermodynamic ``costume" for the uncertainty function $F$ that
sheds some new light onto the meaning of both $\hbar$ and
$\omega$.

\vskip 2mm \nd {\it In particular, we see that $\hbar/2$ is the
derivative of the uncertainty function $F$ with respect to the
delocalization factor $D$.} Increases $dF$ of the thermal
uncertainty function $F$ are of two types  {\sf (our 7th. new
result)}
\begin{itemize} \item i) from the excitation energy, that supplies
a $C/\omega$ contribution and \item ii) from the delocalization
factor~$D$
\end{itemize}

\nd  Additionally, on account of Eq.~(\ref{nuevaincert}), on the
one hand, and since the degree of purity reads
[Cf.~Eq.~(\ref{purito})] $d_{\mu}=I_{\tau}/2 $, on the other one,
we are led to {\it an uncertainty relation for mixed states in
terms of} $d_\mu$, namely,

\be \label{mixta}
\Delta_{\mu}x\,\Delta_{\mu}p=\frac{\hbar}{2}\,\frac{1}{d_{\mu}},
\ee that tells us just how uncertainty grows as participation
ratio $R =1/d_\mu$ augments {\sf (8th. new result)}.
Eq.~[(\ref{mixta})]
 is of semi-classical origin, which makes it a bit different from
 the one that results form a purely quantal treatment
 [see~\cite{dodonov},~Eq.~(4)].

%%%%%%%%%%%%%%%%%%%%%%%%%%%%%%%%%%%%%%%%%%%%%%%%
\section{ Degrees of purity's relations}
%%%%%%%%%%%%%%%%%%%%%%%%%%%%%%%%%%%%%%%%%%%%%%%%%%%
\label{purities1}

\nd We relate now the degree of purity of our thermal state with
various physical quantities both  in its quantal and
semi-classical versions {\sf (our 9th. new result)}. This will
explain the upper bound 1/2 of the semi-classical purity $d_\mu$.
Using Eqs.~(\ref{virial}) and (\ref{mixta}) we get

 \be \label{drelate} d_{\mu}= \frac{I_\tau}{2}= (1-d_{\mu})\,d_{\rho},
 \ee
 which leads to

 \ben \label{drelate2}
  d_{\rho}& =&
 \frac{d_{\mu}}{1-d_{\mu}}=\frac{I_\tau}{2-I_\tau},\nonumber\\
 d_{\mu}& =& \frac{d_{\rho}}{1+d_{\rho}}
 \een
such as clearly shows that (i) $d_{\mu} \le d_{\rho}$, and (ii)
for
 a pure state, its  semi-classical counterpart has a degree of purity equal
 $1/2$. Moreover, {\it notice how information concerning the purely quantal notion of purity
 $d_\rho$ is already contained in the semi-classical measure} $I_\tau$.

 \vskip 2mm \nd    We appreciate the fact that $R$ increases as delocalization
grows, a quite sensible result. Fig.~\ref{F3} depicts $d_\mu(T)$,
a monotonously decreasing function, which tells us that degree
of purity acts here as a thermometer.  Also, from Eq.~(\ref{IF})
we see that $\beta \hbar \omega=-\ln{(1-I_{\tau})}$. Thus, we can
rewrite Eq.~(\ref{thermal2}) in the following form

 \be \label{nova}
W-S = 1+ \beta\hbar\omega \frac{2d_\mu -
1}{2d_\mu}=1+\beta\hbar\omega \frac{d_{\rho}-1}{2 d_{\rho}},\ee
which casts the difference between the quantal and semi-classical
entropies in terms of the degrees of purity. From Eq.~(\ref{nova})
we can also give the
 quantal mean energy in  terms of $d_\mu$ using Eqs.~(\ref{IF})
  and (\ref{Fishtextbook})

\be E=\frac{ \hbar\omega}{2}
\frac{1-2d_\mu}{d_\mu}=\frac{\hbar\omega}{2}\, \frac{1-d_{\rho}}{d_{\rho}}. \label{Edmu}\ee

%%%%%%%%%%%%%%%%%%%%%%%%%%%%%%%%%%%%%%%%%%%%%%%%%%%%%%%%%%%%%%
\section{Generalizations}
\label{HIOplus} Eqs.~(\ref{HOH}) and (\ref{z}) can be generalized.
It is well known (see, for instance,~\cite{delape} and references
therein) that, for more general Hamiltonians $\mathcal{H}_G$ with
discrete spectra, one can always find a representation that writes
\be \label{delapen} H_G= a_0+ a_1 \eta^{\dagger} \eta + a_1^* \eta
\eta^{\dagger}, \,\,(a_0,\,a_1\,\,{\rm given\,\,constants}), \ee
where $\eta$, of eigenvalue  $z_{\eta}$,  is a generalized
annihilation operator that can be obtained using a definite
algorithm~\cite{delape}. The equation $\eta \vert z_{\eta}\rangle
=z_{\eta} \vert z_{\eta}\rangle$ generates generalized coherent
states and, in turn, generalized Husimi distributions. Most of our
results above follow by replacement of $\hbar \omega$ by a
suitable combination of $a_0$, $a_1$, and $a_1^*$. Work in such
direction is currently being performed.

%%%%%%%%%%%%%%%%%%%%%%%%%%%%%%%%%%%%%%%%%%%%%%%%%%%%%%%%%%
\section{Conclusions}
%%%%%%%%%%%%%%%%%%%%%%%%%%%%%%%%%%%%%%%%%%%%%%%%%%%%%%%%%%%
   \label{conclusions}

\nd Our statistical semi-classical study  yielded, we believe,
some new  interesting physics  that we proceed to summarize. We
have, for the HO,
\begin{enumerate}
 \item  established that the semi-classical Fisher measure $I_\tau$ contains all
 relevant statistical quantum information,
\item shown that the Husimi distributions are MaxEnt ones, with
the semi-classical excitation energy $\mathcal{E}$ as the only
constraint, \item  complemented the Lieb bound on the Wehrl
entropy using $I_\tau$, \item seen in detailed fashion how
delocalization becomes the counterpart of energy fluctuations,
\item  written down  the difference $W-S$ between the
semi-classical and quantal entropy also in $I_\tau-$terms, \item
provided a relation between energy excitation and degree of
delocalization, \item  shown that the derivative of twice the
uncertainty function $F(\beta\omega)=\Delta x \Delta p$ with
respect to $I_\tau^{-1}$ is the Planck constant $\hbar$,  \item
established a semi-classical uncertainty relation in terms of the
semi-classical purity $d_\mu$, and \item expressed both $d_\mu$
 and the quantal degree of purity in terms of $I_\tau$.
\end{enumerate}

%%%%%%%%%%%%%%%%%%%%%%%%%%%%%%%%%%%%%%%%

\newpage

%
%  List of Figures
%

\begin{figure}
\caption{Numerical computation results for the HO:  changes
$dI_\tau$ and $dU$ vs. $dS$ that ensue after randomly generating
variations $\delta p_i$ in the underlying microscopic canonical
probabilities $p_i$.} \label{F1}
\end{figure}

\begin{figure}
\nd \caption{$S$, $\beta E$, and $\ln{D}$ as a function of
$t=T/(\hbar \omega$).} \label{F2}
\end{figure}

\begin{figure}
\caption{Semi-classical purity $d_{\mu}$  vs. $T/h\nu$, a
   monotonous function.}  \label{F3}
\end{figure}


\begin{thebibliography}{99}
%%%%%%%%%%%%%%%%%%%%%%%%%%%%%%%%%%%%%%%%%

\bibitem{roybook}  B.R.~Frieden, {\it Physics from Fisher information}
   (Cambridge University Press, Cambridge, England, 1998).
 %%Physics from Fisher information

\bibitem{roybook2}  B.R.~Frieden, {\it Science from Fisher information}
   (Cambridge University Press, Cambridge, England, 2003).

\bibitem{ThWootters} W.K.~Wootters, {\it The acquisition of information from quantum
   measurements}, PhD dissertation, University of Texas at Austin,~1980.

\bibitem{Frieden} B.R.~Frieden,  B.H.~Soffer, {\it Phys.~Rev.~E} {\bf
52}, 2274 (1995).
  %% ''Lafrangians of physics and the game of Fisher-informationtransfer''


  \bibitem{Renyi}  F.~Pennini, A.R.~Plastino, and A.~Plastino, {\it
Physica  A} {\bf 258}, 446 (1998).
 %% "R\`{e}nyi entropies and Fisher informations as measures of nonextensivity
 %% in a Tsallis setting"

 \bibitem{FPPS}  B.R.~Frieden, A.~Plastino, A.R.~Plastino, and
H.~Soffer, {\it Phys. Rev. E} {\bf 60}, 48 (1999).
 %%"Fisher-based thermodynamics: its Legendre transforma and concavity
 %%properties''

 \bibitem{Incerteza} F.~Pennini, A.~Plastino, A.R.~Plastino, and
 M.~Casas, {\it Phys. Lett. A} {\bf 302}, 156 (2002).
  %%''How fundamental is the character of thermal uncertainty relations?''


 \bibitem{wehrl} A.~Wehrl, {\it Rep.~Math.~Phys.} {\bf 16}, 353 (1979).
%%''General properties of entropy''.

\bibitem{Lieb} E.H.~Lieb, {\it Commun.~Math.~Phys.} {\bf 62}, 35 (1978).



\bibitem{Pennini1} F.~Pennini and A.~Plastino, {\it Phys.~Rev.~E} {\bf
69}, 057101 (2004).
%%''Heisenberg--Fisher thermal uncertainty measure''


\bibitem{wigner} E.P.~Wigner, {\it Phys.~Rev.} {\bf 40}, 749 (1932).


\bibitem{Wlodarz} J.J.~Wlodarz, {\it Int.~J.~Theor.~Phys.}  {\bf 42}, 1075 (2003).

\bibitem{husimi}  K.~Husimi, {\it Proc.~Phys.~Math.~Soc.~Japan} {\bf 22}, 264 (1940).

\bibitem{oconnel} R.F.~O'~Connel and E.P.~Wigner,{\it  Phys.~Lett.~A}
{\bf 85}, 121 (1981).


\bibitem{mizrahi} S.S.~Mizrahi, {\it Physica A} {\bf 127}, 241 (1984);
ibidem {\bf 135}, 237 (1986); {\bf 150}, 541 (1988).


\bibitem{PRD2753_93} A.~Anderson and J.J.~Halliwell, {\it
 Phys.~Rev.~D} {\bf 48}, 2753 (1993).
 %%''Information-theoretic measure of uncertainty due to quantum and thermal
%%  fluctuations''

\bibitem{anderson} M.H.~Anderson {\it et al.}, {\it Science} {\bf
269}, 198 (1995).


\bibitem{davis} K.B.~Davis {\it et al.}, {\it Phys. Rev. Lett.} {\bf
75}, 3969 (1995).

\bibitem{bradley} C.C.~Bradley, C.A.~Sackett, and R.G.~Hulet,
 {\it Phys. Rev. Lett.} {\bf 78}, 985 (1997).

\bibitem{cramer} H.~Cramer, {\it Mathematical methods of statistics},
 (Princeton University Press, Princeton, NJ, 1946).

\bibitem{Gnutzmann} S.~Gnuzmann, K.~$\dot{Z}$yczkowski, {\it J.~Phys.
A} {\bf 34}, 10123 (2001).


\bibitem{Glauber} R.J.~Glauber, {\it Phys. Rev.} {\bf 131}, 2766 (1963).
%% ''Coherent and Incoherent States of the Radiation field''.


\bibitem{klauder} J.R.~Klauder and B.S.~Skagerstam, {\it Coherent
states} (World Scientific, Singapore, 1985).


\bibitem{Schnack} J.~Schnack, {\it Europhys. Lett.} {\bf 45}
(1999) 647.
%% ''Thermodynamics of the harmonic oscillator''.

\bibitem{katz}  E.~T.~Jaynes, {\it Phys.~Rev.} {\bf 106}, 620 (1957); {\bf
108}, 171 (1957);  1987 {\it Papers on probability, statistics and
statistical physics, edited by  Rosenkrantz R. D.} (Dordrecht:
Reidel); A.~Katz, {\it Principles of Statistical Mechanics:
The Information Theory Approach} (San Francisco: Freeman and Co.), 1967.


\bibitem{ufano} W.J.~Munro, D.F.V.~James, A.G.~White, and P.G.~Kwiat,
{\it Phys.~Rev.~A}  {\bf 64}, 03030202  (2003);  U.~Fano, {\it
Rev.~Mod.~Phys.} {\bf 29}, 74 (1957).

\bibitem{casas}  J.~Batle, A.R.~Plastino, M.~Casas, and A.~Plastino,
{\it J. Phys. A: Math. Gen.} {\bf 35}, 10311 (2002).

 \bibitem{pathria1993}
 R.K.~Pathria, {\it Statistical Mechanics} (Pergamon Press,
  Exeter, 1993).

\bibitem{dodonov} V.V.~Dodonov, {\it J. Opt. BA} {\bf 4}, S98 (2002).


\bibitem{canosa} N.~Canosa, A.~Plastino, and R.~Rossignoli, {\it Phys.~Rev.~A} {\bf 40}, 519 (1989).

\bibitem{sugita} A.~Sugita and H.~Aiba, {\it Phys. Rev. E} {\bf
65},  032205 (2002).

\bibitem{Pennini2} F.~Pennini and A.~Plastino, {\it Phys. Lett. } A {\bf
 326}, 20 (2004).


\bibitem{mandelbrot} B.~Mandelbrot, {\it Ann. Math. Stat.} {\bf
33}, 1021 (1962).

 \bibitem{flang} F.~Pennini and A.~Plastino, {\it Physica A} {\bf
 334}, 132 (2004).

\bibitem{delape} L.~de~la Pena and R.~Montemayor, {\it Am.~J.~Phys.} {\bf 48}, 855 (1980).




  \end{thebibliography}
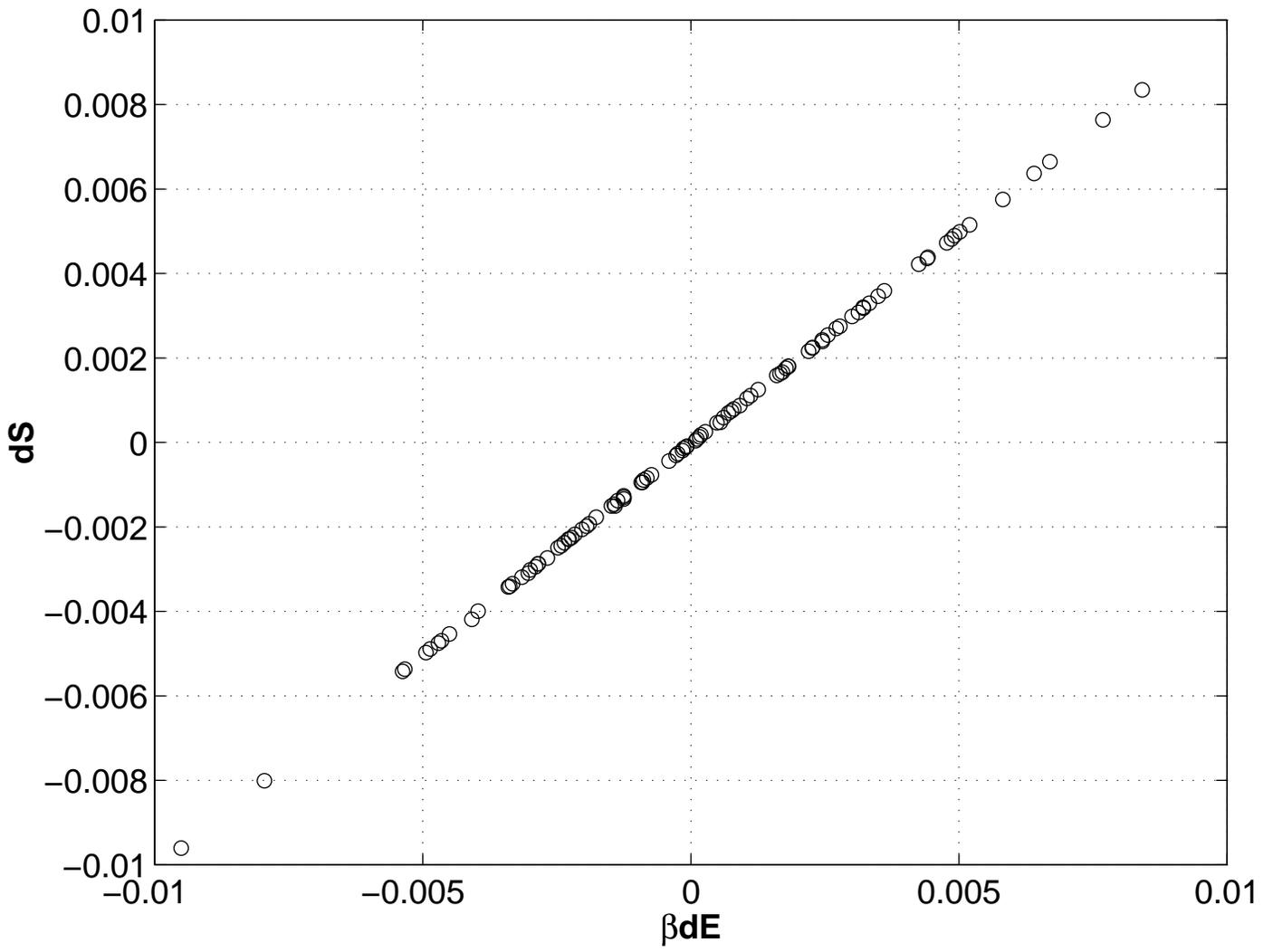
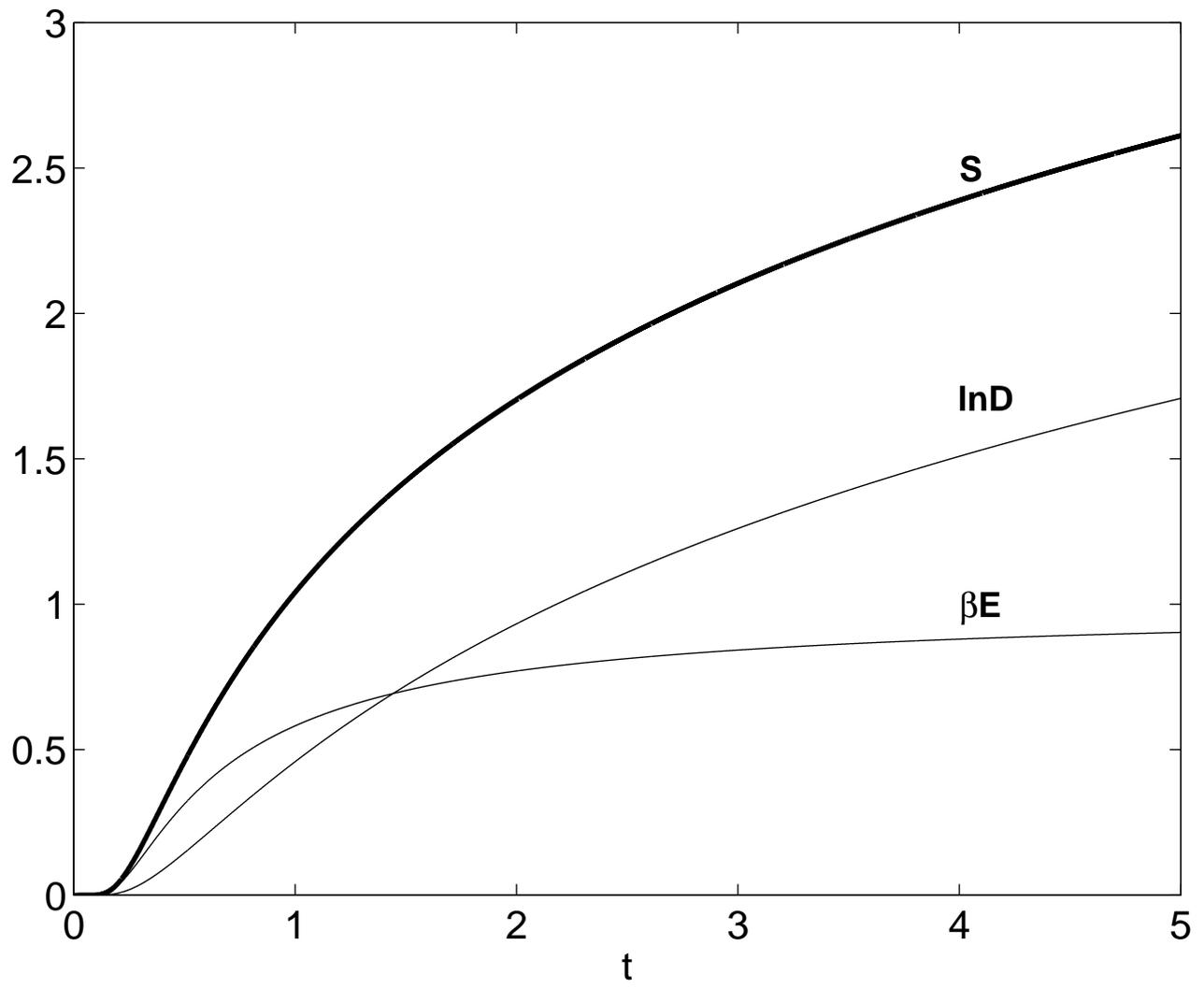
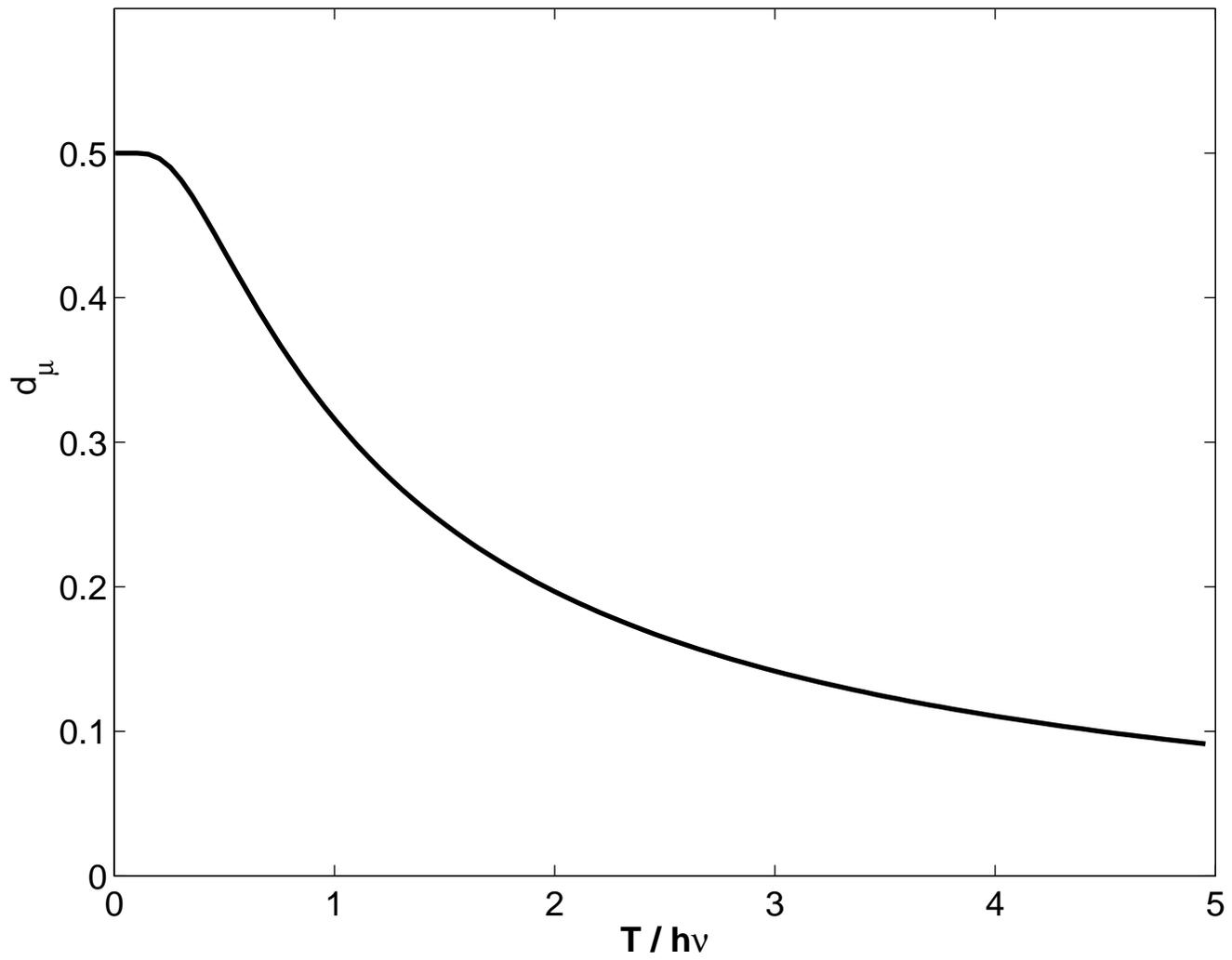
\end{document}